# Finding top performers through email patterns analysis

Wen, Q., Gloor, P. A., Fronzetti Colladon, A., Tickoo, P., & Joshi, T.

# Finding top performers through email patterns analysis

Wen, Q., Gloor, P.A., Fronzetti Colladon, A., Tickoo, P., & Joshi, T

**Abstract:** In the information economy, individuals' work performance is closely associated with their digital communication strategies. This study combines social network and semantic analysis to develop a method to identify top performers based on email communication. By reviewing existing literature, we identified the indicators that quantify email communication into measurable dimensions. To empirically examine the predictive power of the proposed indicators, we collected a 2 million email archive of the 578 executives in an international service company. Panel regression was employed to derive interpretable association between email indicators and top performance. The results suggest that top performers tend to assume central network positions and have high responsiveness to emails. In email contents, top performers use more positive and complex language, with low emotionality, but rich in influential words that are more likely reused by coworkers. To better explore the predictive power of the email indicators, we employed AdaBoost machine learning models, which achieved 83.56% accuracy in identifying top performers. With cluster analysis we further find three categories of top performers, "networkers" with central network positions, "influencers" with influential ideas, and "positivists" with positive sentiments. The findings suggest that top performers have distinctive email communication patterns, laying the foundation for grounding email communication competence in theory. The proposed email analysis method also provides a tool to evaluate the different types of individual communication styles.

**Keywords:** top performer; email communication; social network analysis; semantic analysis.

# 1. Introduction

Effective communication becomes increasingly indispensable to achieve high work performance in an age of hyper-specialization, as there is a need for intensive information sharing to integrate multidisciplinary expertise [1]. The rapid development of electronic communication technology enables people to continuously keep in contact, free from the restrictions of time and space [2, 3]. In order to facilitate effective communication, many firms are making huge investments in advanced information systems, however, without fully understanding the effects of communication patterns on performance [4]. Academic research is especially lacking on how email communication patterns are associated with work performance [5, 6].

Email communication has long been reported to occupy a significant proportion of knowledge workers' time [7, 8]. It provides a ubiquitous information sharing channel, in which people are available even when they are physically absent [9]. Due to its large information carrying capacity, emails have become a dominant means of communication in large organizations. Although emails have high potential to reflect organizational communication patterns and behaviors, they have only been used in a limited number of past studies [7]. This is primarily attributable to two difficulties. Firstly, emails miss some important behavioral cues that are integral to face-to-face communication, such as speech tone and facial expressions [10]. However, some researchers argued that people tend to compensate for this shortcoming with additional cues in emails, such as the use of capitalized words and emoticons [2]. Secondly, email data are inherently unstructured, making it hard to quantify constructs and test research hypotheses [7]. However, recent developments in social network analysis (SNA) and text mining methods enable large-scale unstructured data analyses, which are promising for identifying communication behaviors from emails [11, 12].

Despite the critiques and arguments, it is clear that email data can act as a rich information source, from which meaningful signals of communication behaviors can be extracted [13, 14]. Some pioneering studies have already identified several behavioral indicators of email communication [3, 7, 14], however, in a relatively fragmented manner, with each of them focusing on a few particular indicators. This study aims to further these research efforts and systematically investigate how the rich information in email communication can be operationalized into quantifiable indicators that are predictive of individuals' work performance. Thus, this study addresses the following research question: How can the information in email communication be aggregated to identify top performers in organizations?

Based on the email archive of 578 executives working in a global software services company, we combined regression and machine learning models to examine how top performers can be identified with email indicators. The findings reveal the most predictive email indicators and the various email communication patterns of top performers. By addressing the research question, this study provides contributions in two areas. First, it develops a set of email communication indicators and demonstrates their predictive power in identifying top performers. The identified email communication indicators of top performers lay foundation for operationalizing communication competence in the context of email communication from a social network perspective. Second, the identified influential email indicators provide a practical tool to map different individuals' contributions to organization communication. This facilitates the reflection on individuals' communication patterns and in turn generates valuable implications for improving email communication efficiency.

The rest of the study is organized as follows. The second section reviews the related literature on communication and work performance, and identifies the research gap to be addressed. The third section describes the data collection and analysis procedures, followed by the empirical analysis results in the fourth section. The fifth section discusses the theoretical and practical implications of the findings. Finally, the sixth section draws the conclusions.

## 2 Literature review

### 2.1 Email communication and individual performance

As a basic information channel for modern organizations, emails can reflect not only communication behaviors but also various other behaviors beyond communication [2, 4, 7]. Correspondingly, two streams of research implied that email communication contains meaningful signals that are predictive of individuals' performance. The first contains abundant research efforts to explore how team- and individual-level communication influences performance. Through the theoretical lens of the conduit model of communication, team communication is instrumental to effective teamwork [15]. It serves the purpose of diffusing task-relevant information [16], addressing coordination problems [17] and developing a shared understanding of team states [18, 19]. Many researchers further argued that communication not only reflects but also in turn shapes team states and even organization institutional settings [15, 20], which have profound influence on performance. Following these theoretical arguments, empirical findings substantiated the positive relationship between email communication and team performance (see [21] for a meta-analysis). In this light, effective email communication is indispensable for becoming top performers in teamwork.

At the individual-level, communication competence is one of the most frequently cited enablers of superior work performance [22, 23]. With increasing work virtualization, many studies propose the notion of individuals' "virtual competence" (and similarly "virtual intelligence" in [24]), of which email communication is an important ingredient. The number of communication relationships is closely related to individuals' ability to access information and resources [25, 26]. The diversity of communication relationships is advantageous for individuals when performing tasks that require multidisciplinary knowledge [27]. Lower communication frequency was found to be associated with less voice in team decisions [28], more feelings of isolation [29] and lower work performance [30].

On the other hand, some researchers pointed to the undesirable effects of frequent email communication that emails may act as a potential source of distraction and work stress [31]. Dealing with emails and recovering from the interruption caused by emails were reported to consume a considerable proportion of knowledge workers' time [32]. The pressure of responding to a number of emails can easily increase individuals' work stress [33]. These effects reduce individuals' work performance and contribute to the productivity puzzle of "being able to do more work but not to do work more productively" in information society [34]. Many research efforts have been devoted to empirically examine the effects of email communication on work performance, but the findings are mixed and the effect of email communication is still controversial.

Despite the debates on how email communication influences performance, existing studies suggest that the quality, diversity and frequency of communication are influential to work performance, and hence imply a large potential of emails in predicting performance [35, 36]. As pointed out by Barley et al. [33], the conflicts in empirical findings are partly attributable to the fact that most existing research measured

email communication with questionnaire instruments instead of real-world email data. This is echoed by Aral et al. [36], who utilized email data as an objective measure of communication that contributes to overcoming the bias in surveys based on participants' memory of their communication networks. In this light, we expect that email communication is predictive of work performance and empirically examine this possibility.

The second stream of closely related research focuses on the behavioral cues embodied in email communication. The large information carrying capacity of email makes it a rich source of individual-level interactions [7–9, 12]. According to social attribution theory, people attribute communication relationships to be not only *instrumental* to fulfill organizational roles and obligations, but also *expressive*, for example, to voice feelings and gain satisfaction [37, 38]. Therefore, email can reflect various individual behaviors, and in fact, some pioneering research already identified multiple kinds of behaviors from email data. For example, Fragale et al. [7] utilized email to measure deference behaviors and argued that email data are advantageous for studying behaviors that are easily distorted by researchers' interventions. The studies by Mazmanian et al. [9], Mazmanian et al. [8] and Butts et al. [3] all suggest that email contains the information on work-nonwork conflict and work engagement escalation. Byron and Baldridge's [2] experiment revealed that expression methods (using capitalization and emoticons) of emails have significant influence on the email recipients' impressions of senders' likability. Similarly, Lim and Teo [39] and [40] identified incivility behavior from emails and analyzed its relationship with work-load and work attitude. Although these studies focus on specific behaviors and hence are fragmented, they imply that email contains abundant behavioral cues, which may have profound implications for individual work performance [7, 39]. Extending these findings, we aim to explore how the abundant information in email can be integrated to provide insights for work performance, and particularly, identifying top performers.

## 2.2 Email communication indicators

Compared to traditional questionnaire surveys, email archives provide more authentic recordings of real-world communication behaviors free from retrospective bias [41, 42]. Several studies e.g. [14, 43] developed a system of email indicators, known as "honest signals", to quantify email communication behaviors. However, many more studies explored similar email communication indicators without explicitly referring to the notion.

To develop a set of email indicators based on a comprehensive review of the extant literature, we conducted extensive literature searches in major databases, including INFORMs, Web of Science, Elsevier ScienceDirect and ProQuest. Using "email/e-mail network", "email/e-mail communication" and "electronic communication" as key words, we found 61 studies related to email communication. Based on the 61 studies, we further identified 8 studies relevant to this topic that are cited by them or citing them. This procedure was repeated until no more relevant reference could be found [44]. As an additional validation, we compared the retrieved studies with the reference lists of two relevant meta-analyses [21, 45] and found no other relevant study to be supplemented. In this way, totally 73 studies were retrieved, among which 18 studies developed indicators for email data in real-world work settings. Consistent with the theoretical implications in 2.1, most of the indicators in email communication were found to be significantly associated with information sharing, collaboration and various other outcome variables. We construct a Sankey diagram to illustrate which *study* analyzed which *dimension* using which *indicator* to explain which *outcome variable* (see the Appendix Table 1 for a detailed list).

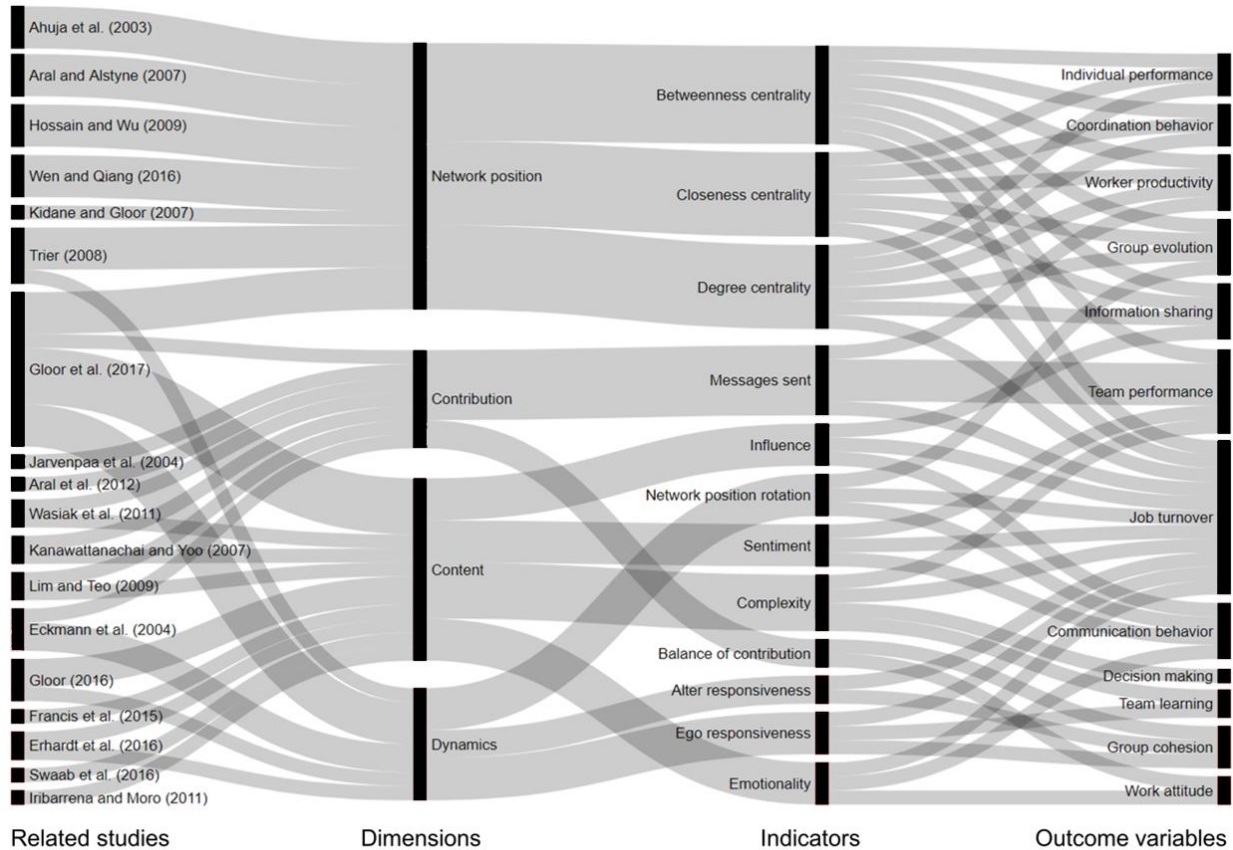

**Figure 1 The Sankey diagram of existing studies using email communication indicators**

A one-unit width of the "flow" in the Sankey diagram corresponds to one email indicator adopted in one previous study. In this way, the size of the rectangle corresponding to each study is proportional to the number of email indicators used in that study, and the width of the flow from each study to each dimension is proportional to the number of indicators that study used in that dimension. The size of each indicator reflects how many times the indicator was used in previous studies, and the width of the flow from each indicator to each outcome variable reflects the number of times that indicator was used to predict that outcome variable.

As shown in Figure 1, email network position is the most frequently studied dimension, with the three centrality indicators receiving similar research attention. The contribution dimension attracted much fewer research efforts, the majority of which were devoted to analyzing the number of emails sent. A wide range of studies investigated email content indicators with similar emphasis on the four indicators. The dynamics of email communication is less frequently studied, especially the responsiveness of co-workers (alter responsiveness).

The links between indicators and outcome variables are rather dispersed. This reflects the fact that there is a lack of wide consensus on which indicator has particularly good predictive power over an outcome variable, and many studies are still exploring using a variety of indicators e.g. [13]. Only a few studies explicitly examined individuals' performance using email communication indicators, they primarily focused on the effects of network position [41, 42]. This coincides with the argument that most previous studies on information advantages in communication networks are "content agnostic" [27]. Therefore, existing studies indicate the potential of email indicators in predicting individuals' work performance, whereas how the indicators can be used to identify top performers still remains to be explored.

## 3. Methods

In order to empirically explore how the email indicators can be used to predict individual performance, we designed an empirical analysis procedure as illustrated in Figure 2. In step one, the data is collected from the company's e-mail server at the end of each month for the duration of six months. For privacy reasons the collection process is executed on the company's server. In step two, using the dynamic semantic social network analysis software Condor [46], the different social network metrics described in the remainder of this section are calculated twice, at the beginning and the end of the observation period, and exported as a table. In step three, regression analysis is applied to obtain interpretable association between each email indicator and individual work performance. In step four, machine learning is employed to further explore the predictive power of the email indicators.

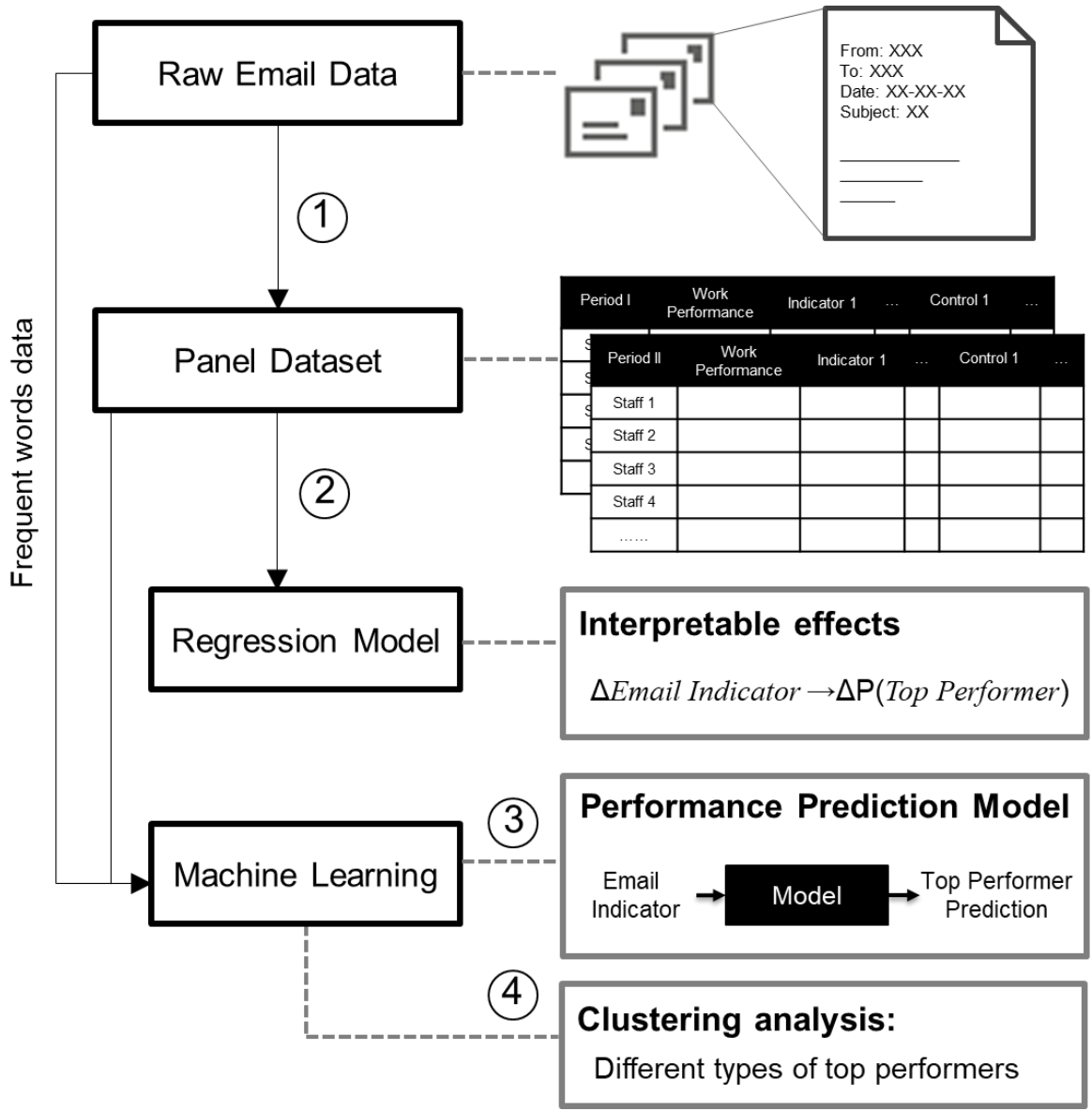

**Figure 2 The explorative empirical analysis procedure**

## 3.1 Data collection and indicator calculation

We collected email data from the top 578 executives of a global services company with over 70,000 employees. The email dataset is ideal for the purpose of this study since the 578 executives coordinate the company's global operation and use email as the primary communication channel. With the help of the software tool Condor [46], we were able to calculate email communication indicators from raw email data on the company's server without directly reading email contents. This approach protects staffs' privacy and, at the same time, collects the data ready for further analysis. Corresponding to the two sets of performance rating data, we included two waves of email indicator data calculated from the more than 2 million emails between the 578 executives and other staff members into our analysis, at the beginning and the end of the observation period. To take full account of intra-organization email communication relationships, we set the boundary of the email network as the whole company and derived the network indicators based on the whole network. The details on indicator calculation methods are described in the next paragraph.

### 3.1.1 Performance rating as the dependent variable

Executives' performance was assessed and rated by their supervisors working together with human resource managers. Executives were assessed using the standard performance evaluation process of the company, based on achievement of their individual goals agreed with their supervisors, and an assessment of their "growth potential" based on feedback from peers, supervisors, HR, and people reporting to the executive. The ratings were binary, indicating whether each executive was a top performer in that period or not. Executives met regularly with their supervisors and HR managers, to appraise their past performance and set goals for the future. Individual performances were compared with those of all executives in the company in order to classify them as top performers or not. Final evaluations also impacted bonuses and promotions.

We analyzed the entire population of the 578 executives in the company, i.e. all their email accounts. We obtained two sets of performance ratings in Jan-Feb 2017 and Apr-May 2017 from the company.

### 3.1.2 Network position indicators

We constructed a directed graph where nodes represent executives' email accounts, with an arc originating at node A and terminating at node B, if A sent at least one email to B. Arcs were weighted according to the number of emails sent from one node to another.

Betweenness, closeness and degree centrality indicators are among the most widely used indicators of network position in email communication analysis [47]. For example, centrality in email communication networks has been found to influence productivity [42], predict job turnover [13] and mediate the relationship between formal position and performance [41]. Thus, we expect that higher centrality in email communication network is associated with superior work performance. The three centrality indicators were calculated with Condor using the following formula [48, 49] where $\#d_{ij}$ is the number of shortest email communication paths from $i$ to $j$ and $\#d_{ij}(t)$ is the number of those paths that pass through $t$; $d_{ij}$ is the distance from $i$ to $j$. N is the number of nodes in the graph.

$$Betweenness\ Centrality = \sum_{\{i,j \neq t\}} \frac{\#d_{ij}(t)}{\#d_{ij}}$$

Betweenness centrality corresponds to the probability of being on the shortest path in the network and is commonly taken as a proxy for power and influence of a person in the network.

$$Closeness\ Centrality = \frac{N-1}{\sum_j d_{ij}}$$

Closeness centrality describes the average number of steps one has to take to reach any other person in the network and is a proxy for the embeddedness of a person in the network.

$$Degree\ Centrality = Number\ of\ email\ communication\ relationships$$

Degree centrality describes the number of direct links of a node, i.e. the number of people a person sends emails to, or receives emails from. This metric can be taken as a proxy for information diversity that a person is exposed to.

### 3.1.3 Network contribution indicators

The amount of information contributed by an individual to the whole network can be approximated by the number of emails sent by the individual. Some researchers suggest that email content should also be considered to eliminate non-work-related emails [41, 50]. However, human coding would require a huge effort for our dataset. So we adopted the total number of emails sent as a proxy herein and address email contents with content indicators and automatic content analysis in the machine learning part.

Besides the absolute contribution, how much an individual contributes information compared to how much s/he receives information from others is also an important relative indicator [4, 27, 39]. An intuitive indicator of relative information contribution is the Contribution Index [46]:

$$Contribution\ Index = \frac{Number\ of\ message\ sent - Number\ of\ messages\ received}{Number\ of\ message\ sent + Number\ of\ messages\ received}$$

### 3.1.4 Network dynamic indicators

Email communication is an inherently dynamic process, in which individual network positions may be continuously changing over time. At the team-level, existing studies suggest that oscillation in network

positions enables leadership rotation and is beneficial for mobilizing diverse participants' advantages over time [51, 52]. However, it remains to be examined whether these findings can be generalized to the individual level. Following Gloor et al. [46], we specifically focus on the oscillation in betweenness centrality and calculate it as the number of times that an individual changed her/his role from local maxima or minima back and forth on a weekly basis [53].

Another important email communication dynamics is responsiveness. Reciprocity is the premise of effective communication [7, 37], especially for asynchronous email communication [54]. Some previous studies found that higher responsiveness represents a positive signal of respect and is associated with superior team learning [4] and lower job turnover tendency [13]. However, other researchers argued that promptly responding to emails can cause distractions [32], increase stress level [31, 33] and hence negatively influences individuals' performance [55]. Therefore, there is a need to empirically examine how an individual's work performance is influenced by his/her own and his/her co-workers' responsiveness. Specifically, we consider the average response time (ART) and the "nudges", defined as the average number emails sent from one person to another before getting a response [13]. For each individual, *ego ART* is the average time needed for the individual to respond, *ego nudges* is the average number of emails sent to the same recipient before getting a response, *alter ART* is the average time the co-workers take to respond to the individual, and *alter nudges* is the average number of emails received from an individual from his/her co-workers before he sends a new email to them. As well as the other metrics we mentioned, both nudges and ART were calculated using the software Condor.

### 3.1.5. Network content indicators

Due to privacy restrictions we were not able to directly read the full emails. For the purpose of this analysis, we relied on the content indicator calculation program in Condor to obtain the content indicator data. The sentiment of email content reflects the sender's mood state and also potentially influences the receiver's mood state [56]. Some empirical evidences suggest that email sentiment is predictive of individual job turnover [13, 57] and team performance [18]. There are many email sentiment calculation approaches, including manual rating, lexicon-based methods or machine learning models. The Condor software provides a built-in sentiment analysis function based on a Bayesian classifier, which has been trained on billions of tweets and achieves over 80% accuracy on many English email corpora [58]. We directly utilized this function to get the sentiment score of each email varying from 0-negative to 1-positive and then averaged over each individual's emails to derive a sentiment score that reflects her/his average sentiment level. Therefore, average sentiment is a score obtained for each individual, which was subsequently used to see whether each executive had a language more or less positive than his/her peers. We also calculated the deviation from neutral sentiment of each individual's emails as the emotionality indicator [13]. The idea behind this indicator is that a language that contains less neutral, more strongly positive or negative, expressions is more emotional.

The informativeness of the email content is another frequently studied aspect [27]. The complexity indicator is calculated based on the likelihood distribution of words within an email, i.e. the probability of each word to appear in the text based on the term frequency/inverse document frequency (TF-IDF) information retrieval metric [58].

$$Complexity\ index = \frac{1}{n}\sum_{w \in V} q(w) \log \frac{1}{p(w)}$$

where $n$ is the total number of words within an email, $V$ is the vocabulary of words that appear in the email corpus, $q(w)$ is the frequency of word $w$ in the email, $p(w)$ is the probability of word $w$ to appear in an email and $\log \frac{1}{p(w)}$ is the inverse document frequency of word $w$ in the corpus. TF-IDF has also other variants and weighting schemes, different from the one used in this research e.g. [59–61]. We chose to adopt this version as it is the one implemented in the Condor software and it led to significant results in past research e.g. [13, 14, 52].

The complexity indicator is the opposite of the log-likelihood of the email text. Therefore, it measures the extent to which an email uses rare (complex) words and introduces non-redundant information. The measure is then averaged at the individual level. It can also be regarded as a word-level analogy to the previous measure of information diversity in emails [27].

Beside the complexity of an email itself, the extent to which co-workers will adopt new ideas and reuse the words that identify them is calculated as the influence indicator. Each time a receiver receives an

email, his/her subsequent emails - sent within four days[1] - are retrieved and combined to derive a word distribution (i.e. how many times each word appears in the emails content). This word distribution is transformed into a vector of the dimensionality of vocabulary ($V$) using TF-IDF and is compared with the TF-IDF vector of the original email based on the cosine similarity measure, which is widely used in text mining [62]:

$$Influence\ index = cos(\frac{\sum_{w \in V} TI_s(w)TI_r(w)}{\sqrt{\sum_{w \in V} TI_s(w)^2} \sqrt{\sum_{w \in V} TI_r(w)^2}})$$

where $TI_s(w)$ is the TF-IDF value of word $w$ in the sender's original email, $TI_r(w)$ is the TF-IDF value of word $w$ in the receiver's subsequent emails.

If an individual's unique words (with high TF-IDF values) were adopted by co-workers in subsequent emails, we can expect that the idea expressed by these words was influential and being spread across the network [63]. However, the timing of emails is the only mechanism controlling the direction of influence. There may be some confounding factors (e.g. common experience or face-to-face communication) that randomly cause co-workers to use the same words. As we averaged the influence score of all the emails for each individual, we assume that such error will cancel out in the large email dataset. On the other hand, we acknowledge this as a limitation, concerning which the findings on the influence index should be interpreted with caution.

### 3.1.6 Control variables

We collected basic personal information from the company's human resource system to construct control variables. The collected information was matched with other variables extracted from the company's HR database. Specifically, we controlled the effects of age, formal position (*band*, a binary variable indicating whether an individual holds a higher level formal position or not), tenure in the company and length of time since last promotion (measured in months). These were all the variables that the company was willing to share based on their privacy policy. These variables potentially influence both individuals' email communication behaviors and work performance. So we included them in the regression model to eliminate alternative interpretation of the effects of email indicators. Table 1 summarizes the variables that we described in this and previous sections.

---

[1] The number of days was determined by trial and error by the algorithm developer to find the influence indicator with the strongest explanatory power (Brönnimann, 2014).

### Table 1 Summary of Study Variables

| Variable | Short Description |
|---|---|
| Betweenness | A centrality metric, which indicates how many times a social actor is on the shortest path to all other actors. |
| Closeness | Measures the average number of steps a person has to take to reach all his/her peers in the network. |
| Degree | Counts the number of direct email contacts of an individual, i.e. the total number of people from/to whom s/he sent/received an email. |
| **Betweenness Oscillations** | Counts the number of times an individual changes her/his role in terms of betweenness centrality, from local maxima or minima back and forth, averaged over a one-week time window. |
| Messages Sent | Number of emails sent by a person. |
| Balance in Contribution (BOC) | Measured through the Contribution Index, it expresses how many messages are sent by an individual compared to how many s/he receives from others. |
| Influence | Measures the extent to which the words and ideas introduced by a person will be reused by his/her co-workers. |
| Sentiment | Measures the positivity or negativity of the language used, varying in the range [0,1] – where values close to 0 indicate the prevalence of negative words, and a score of 1 a totally positive language. |
| Complexity | Measures the extent to which a person uses rare (complex) words and introduces non-redundant information. |
| Emotionality | Measure the deviation of the language from neutral sentiment, i.e. how intensively negative or positive the words are used in the email content. |
| Nudges | Count of the number of emails sent from one actor to his/her peers before getting an answer (Ego Nudges), or of the number of emails received by peers before the actor responds to them (Alter Nudges). |
| Average Response Time (ART) | Measures the average time taken by a social actor to answer to an email s/he receives (Ego ART). Alter ART measures the average time taken by all recipients of the actor's emails to answer her/him. |
| Age | Corresponds to the age of an individual, expressed in years. |
| Band | A discretionary variable indicating whether an individual holds a higher level executive position. |
| Tenure | Tenure of the person in the company (expressed in months). |
| Time Since Last Promotion (TSLP) | Time which has passed since a person was promoted (expressed in months). |

Note. Complete descriptions of variables are provided in the text.

## 3.2 Partial least square regression model

As widely reported in many previous studies, there exist strong correlations among email network indicators, especially the network centrality indicators [64], raising concerns of a multi-collinearity problem.

Some previous studies directly conducted OLS regression and interpret the result based on the indicators that appear to be significant e.g. [65]. The regression coefficients obtained by simple regression were lacking stability and robustness. Many studies avoid the collinearity among network indicators by retaining only one network indicator that is most theoretically interpretable [64, 66]. However, in view of the explorative nature of this study, a well-established theoretical foundation is not in place to help decide which indicator should be chosen.

To empirically explore the effects of different network indicators, one stream of studies select the indicator that generates the best statistical result. For example, Powell et al. [67] employed stepwise regression to determine which variable to enter in the final model. Kao et al. [68] utilized the feature clustering method that groups strongly correlated variables into clusters and chose one representative variable from each cluster. Gloor et al. [13] explored the effects of three centrality indicators in three separate models to choose the best model based on the information criterion. However, the presence of strong correlation effects may be attributable to the underlying links among indicators, which is not captured when selecting any single representative indicator. From a network theoretical lens, Kilduff and Tsai [69] also recommended to control the effects of the other centrality indicators when analyzing a particular centrality indicator. Another stream of studies propose to introduce latent variables that model the underlying links. For example, de Andrade and Rêgo [70] extracted principal components from network centrality indicators for subsequent analysis. Ahuja et al. [41] further suggest to use a component-based estimation strategy to incorporate the strong correlations among network centrality indicators, and adopted partial least square (PLS) regression to examine the effects of network centrality on work performance.

Combining the insights from these two streams of research, we explored the effects of email communication indicators with generalized PLS regression, which is suitable for our research purpose for two reasons. First, PLS regression addresses the multi-collinearity problem by grouping independent variables into latent components that capture the underlying links among independent variables. Second, it discovers component structure by maximizing model explanatory power rather than putting restrictions on the way independent variables are combined before running the regression. So it serves the explorative purpose of this study. To implement generalized PLS regression with binary dependent variable, we utilized the *plsRglm* package in R. To further address the concerns about multi-collinearity, we also followed Gilsing et al. [71] to estimate multiple sub-models and assess the robustness of model coefficients. Besides, previous studies recommend to use multilevel data to alleviate the multi-collinearity problem of network indicators [67], so we performed regression analyses using two-wave panel data.

## 3.3 Machine learning models

In order to fully explore the predictive power of email communication indicators beyond linear regression and incorporate unstructured data into the prediction, we also employed machine learning models that allow more complex interactions between independent variables.

Email contents are not available to the researchers to protect privacy. As a compromise, we extracted the top 10 words that appeared in each individual's emails from the company's server using Condor's influence algorithm described in section 3.1. These keywords enable a high-level understanding of the topics that each individual talks about most frequently [27]. A limitation to this approach could lie in the fact that some words could have high frequency in a short time frame, but not be among the top 10 in the whole period – therefore, extemporaneous topics might have been excluded from the analysis. This is a possibility which we intend to explore in future research and which we could not address in this study, as privacy agreements did not allow to access the full email content.

We utilized latent Dirichlet allocation (LDA) to extract content features from this unstructured text data. LDA is a probabilistic model that estimates the probability distribution ($P(\vec{T}|D)$) of documents (**D**) over topics ($\vec{T}$), which are defined as probability distributions over words ($P(\vec{w}|T_i)$ for each topic *i*). It has been found to be effective in identifying major topics from various kinds of text data, including emails [12]. For our email keyword dataset, we treated each individual in one period as a unit of analysis (document **D**) and estimated its topic distribution ($P(\vec{T}|D)$) to reflect how much the individual talked about each topic in that period. The meanings of the identified topics can be inferred based on their word distributions as shown in Figure 3. In this way, we transformed the raw keywords into 6 content features.

Combining the email indicators, the control variables and the content features, machine learning models can be trained to identify top performers. Extending the above logistic regression for the binary performance variable, we trained a logistic regression-based Adaboost model that considers more complex interaction among independent variables and uses a resampling strategy to reduce prediction errors [72]. The model parameters were tuned with cross-validation method to select the model with the strongest predictive power. We also performed cluster analysis to explore different types of top performers.

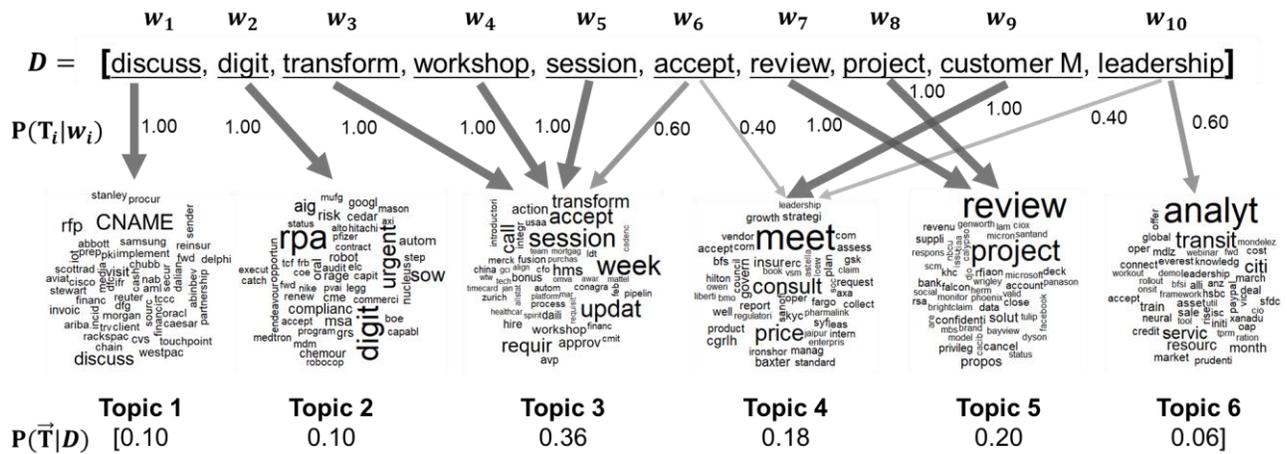

**Figure 3 An illustrative example of the text feature derivation process based on the top 10 words**
**Note**: The words and values in the figure are just indicative and do not reflect the word usage behavior of any specific individual. CNAME represents the company's name.

## 4. Results

The summary statistics and correlations of the variables are presented in Table 2. Many email indicators have strong correlations with the dependent variable (*top performer*), indicating potential predictive power that remains to be further explored.

Table 2 Summary statistics and correlations

|  | min | max | mean | SD | 1 | 2 | 3 | 4 | 5 | 6 | 7 | 8 | 9 | 10 |
|---|---|---|---|---|---|---|---|---|---|---|---|---|---|---|
| 1. Betweenness | 300.80 | 826181.04 | 35022.59 | 51562.03 | - | | | | | | | | | |
| 2. Closeness | 0.27 | 0.50 | 0.37 | 0.01 | 0.25*** | - | | | | | | | | |
| 3. Degree | 28 | 924 | 232.10 | 120.05 | 0.75*** | 0.28*** | - | | | | | | | |
| 4. Messages sent | 15 | 8064 | 1159 | 1142.93 | 0.17*** | 0.59*** | 0.22*** | - | | | | | | |
| 5. BOC | -0.91 | 0.89 | -0.23 | 0.27 | -0.14*** | 0.38*** | -0.19*** | 0.26*** | - | | | | | |
| 6. Influence | 0.21 | 0.73 | 0.18 | 0.33 | 0.19*** | 0.12*** | 0.31*** | 0.21*** | 0.03 | - | | | | |
| 7. Sentiment | 0.28 | 0.81 | 0.63 | 0.04 | -0.15*** | 0.03 | -0.18*** | 0.03 | 0.09*** | -0.19*** | - | | | |
| 8. Complexity | 7.17 | 10.11 | 8.26 | 0.32 | 0.01 | 0.04 | 0.02 | 0.07* | -0.16*** | 0.38*** | -0.13*** | - | | |
| 9. Emotionality | 0.18 | 0.36 | 0.27 | 0.02 | 0.20*** | -0.01 | 0.24*** | 0.00 | -0.15*** | 0.10*** | -0.11*** | 0.15*** | - | |
| 10. Betweenness Oscillations | 14 | 37 | 25.83 | 3.65 | -0.07* | 0.02 | -0.07* | -0.04 | 0.05 | -0.02 | 0.08* | -0.04 | -0.05 | - |
| 11. Alter nudges | 1 | 6 | 1.53 | 0.31 | 0.06 | 0.13*** | 0.04 | 0.08* | -0.52*** | -0.19*** | 0.00 | 0.09*** | 0.10*** | -0.03 |
| 12. Alter ART | 1.92 | 63.45 | 21.42 | 6.66 | 0.14*** | 0.26*** | 0.13*** | 0.12*** | -0.36*** | -0.28*** | 0.04 | 0.07* | 0.14*** | 0.01 |
| 13. Ego nudges | 1.04 | 3.52 | 1.47 | 0.02 | 0.01 | 0.09*** | -0.02 | 0.14*** | 0.59*** | 0.27*** | -0.24*** | 0.03 | -0.03 | 0.04 |
| 14. Ego ART | 6.66 | 40.10 | 20.76 | 5.42 | -0.02 | 0.00 | 0.02 | 0.01 | 0.01 | -0.05 | 0.03 | -0.04 | 0.00 | -0.01 |
| 15. Age | 33 | 67 | 44.82 | 6.13 | 0.41*** | 0.14*** | 0.57*** | 0.09*** | 0.08 | 0.17*** | -0.18*** | -0.07** | 0.14*** | -0.02 |
| 16. Band | 0 | 1 | 0.26 | 0.44 | -0.02 | -0.03 | -0.04 | -0.01 | -0.04 | 0.01 | -0.03 | 0.03 | -0.05 | -0.01 |
| 17. Tenure | 4 | 323 | 99.75 | 71.86 | 0.34*** | 0.10*** | 0.39*** | 0.11*** | -0.26*** | 0.02 | -0.05 | 0.05 | 0.19*** | -0.08** |
| 18. TSLP | 0 | 132 | 30.84 | 25.08 | -0.29*** | -0.21*** | -0.23*** | -0.17*** | 0.11*** | -0.23*** | 0.11*** | -0.03 | -0.13*** | 0.07 |
| 19. Top performer | 0 | 1 | 0.35 | 0.48 | 0.21*** | 0.49*** | 0.23*** | 0.26*** | -0.03 | 0.32*** | 0.26*** | 0.17*** | -0.13*** | 0.01 |

Table 2 Summary statistics and correlations (continued)

|  | 11 | 12 | 13 | 14 | 15 | 16 | 17 | 18 |
|---|---|---|---|---|---|---|---|---|
| 12. Alter ART | 0.66*** | - | | | | | | |
| 13. Ego nudges | -0.43*** | -0.36*** | - | | | | | |
| 14. Ego ART | -0.02 | -0.01 | 0.02 | - | | | | |
| 15. Age | -0.05 | 0.03 | 0.08** | 0.03 | - | | | |
| 16. Band | 0.04 | 0.00 | 0.02 | -0.01 | -0.02 | - | | |
| 17. Tenure | 0.16*** | 0.28*** | -0.09*** | 0.03 | 0.19*** | 0.01 | - | |
| 18. TSLP | -0.05 | -0.04 | 0.02 | -0.01 | -0.40*** | 0.03 | -0.03 | - |
| 19. Top performer | 0.12*** | 0.15*** | -0.27*** | -0.04 | 0.14*** | -0.03 | 0.08** | -0.16*** |

* $p<0.05$; ** $p<0.01$; *** $p<0.001$.

## 4.1 Regression analysis results

First, we estimated a model with only control variables as the baseline. As shown in Table 2, all the four control variables appear to be significantly related to work performance. Second, we ran a regression for each group of email indicators separately with control variables included. Third, a full model including all variables was estimated to compare with the sub-models in the second step and assess the robustness of the model coefficients. For each model, we performed Wald tests on whether the period fixed effects are significant and whether each model coefficient varies significantly across the two periods. If the model coefficients are invariant across periods and the period fixed effect is insignificant, the model is essentially equivalent to a pooled regression. If the model coefficients are invariant but the period effect is significant, a fixed effect model is estimated. If both the model coefficients and the period effect vary significantly across periods, we estimate a variable coefficient model.

The comparison between the full and the sub-models suggests that the majority of model coefficients are consistent and robust, and the model coefficients are visualized in Figure 4. The communication network centrality indicators are all positively related to top performance with *closeness centrality* having the strongest relationship. The coefficients of contribution indicators vary significantly between the full model and the sub-model and across the two periods, indicating that the effects of contribution indicators are not robust. As for content indicators, *influence*, *sentiment* and *complexity* indicators all have significant positive relations with top performance, while *emotionality* is negatively related to top performance. Among the dynamic indicators, only *ego nudges* has a consistently significant relation with top performance, and lower ego responsiveness (higher *ego nudges*) is associated with a lower probability of being a top performer.

**Table 1 Regression analysis results**

|  | Baseline | Position | Contribution | Content | Dynamics | Full |
|---|---|---|---|---|---|---|
| **Position** | | | | | | |
| Betweenness | | 0.028(0.005)*** | | | | 0.045(0.008)*** |
| Closeness | | 0.124(0.009)*** | | | | 0.234(0.038)*** |
| Degree | | 0.032(0.004)*** | | | | 0.030(0.010)** |
| **Contribution** | | | | | | |
| Message sent$_1$ | | | 0.185(0.046)*** | | | -0.013(0.021) |
| Message sent$_2$ | | | -0.189(0.039)*** | | | -0.028(0.021) |
| BOC$_1$ | | | -0.101(0.014)*** | | | -0.022(0.010)* |
| BOC$_2$ | | | -0.036(0.019) | | | 0.011(0.021) |
| **Content** | | | | | | |
| Influence | | | | 0.135(0.010)*** | | 0.180(0.014)*** |
| Sentiment | | | | 0.141(0.016)*** | | 0.109(0.008)*** |
| Complexity | | | | 0.057(0.006)*** | | 0.051(0.008)*** |
| Emotionality | | | | -0.068(0.007)*** | | -0.087(0.011)*** |
| **Dynamics** | | | | | | |
| BetOsc$_1$ | | | | | 0.001(0.002) | 0.003(0.004) |
| BetOsc$_2$ | | | | | 0.008(0.002)*** | 0.008(0.004)* |
| Alter nudges | | | | | -0.034(0.015)* | -0.023(0.014) |
| Alter ART | | | | | -0.037(0.017)* | -0.005(0.009) |
| Ego nudges | | | | | -0.169(0.020)*** | -0.174(0.029)*** |
| Ego ART$_1$ | | | | | -0.045(0.009)*** | -0.021(0.005)*** |
| Ego ART$_2$ | | | | | 0.053(0.017)*** | 0.028(0.006)*** |
| **Controls** | | | | | | |
| Age | 0.019(0.004)*** | -0.009(0.006) | 0.033(0.006)*** | 0.026(0.002)*** | 0.036(0.003)*** | 0.026(0.008)** |
| Band | -0.010(0.003)** | -0.005(0.001)*** | -0.011(0.001)*** | -0.011(0.003)*** | -0.002(0.006) | 0.004(0.004) |
| Tenure | 0.037(0.013)** | 0.008(0.002)*** | 0.005(0.013) | 0.046(0.008)*** | 0.038(0.010)*** | -0.014(0.008) |
| TSLP | -0.073(0.008)*** | -0.018(0.004)*** | -0.053(0.010)*** | -0.070(0.004)*** | -0.065(0.013)*** | 0.008(0.008) |
| **AIC** | 1218.757 | 1189.940 | 1139.242 | 925.529 | 1063.884 | 796.536 |
| **Pseudo R$^2$** | 0.182 | 0.200 | 0.233 | 0.377 | 0.280 | 0.452 |
| **Components** | 2 | 3 | 4 | 3 | 6 | 11 |
| **Model type** | Fixed | Fixed | Variable | Fixed | Variable | Variable |
| **N** | 1156 | 1156 | 1156 | 1156 | 1156 | 1156 |

**Note**: BOC, BetOsc, ART and TSLP stand for balance of contribution, betweenness centrality oscillation, average response time and time since last promotion respectively. Variables that have coefficients significantly different in the two periods are listed with subscripts indicating the period. N is equal to 1156 – which is two times the number of executives in the company (578) – as PLS regression comprised two time periods, i.e. two observations for each person.

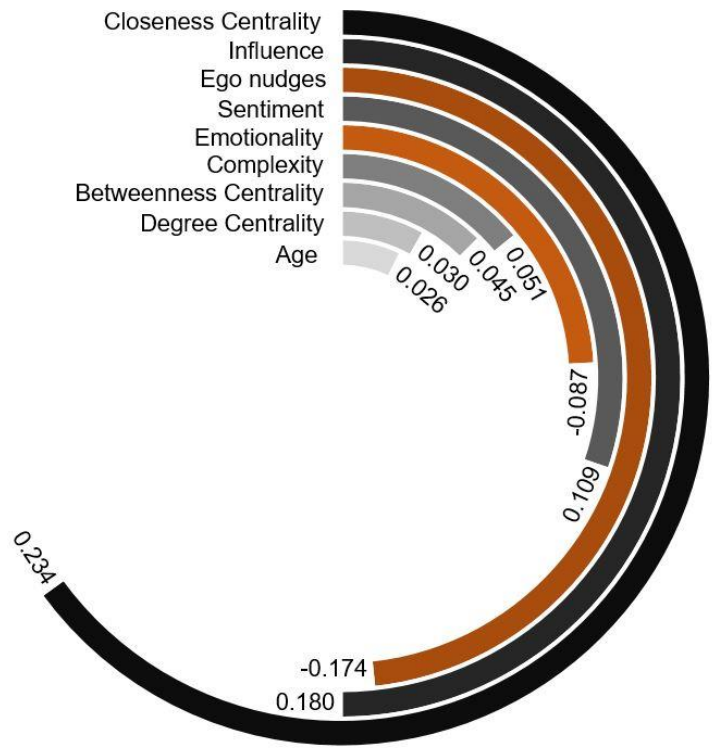

**Figure 4 Significant coefficients in the full model**

## 4.2 Performance prediction models

To better exploit the predictive power of email indicators and allow nonlinear interactions among them, we trained performance prediction models using logistic regression-based Adaboost algorithms. The extracted content features were also included in model training to explore their predictive power.

Logistic regression-based Adaboost models were trained using cross-validation to select the best model. The model accuracy was evaluated using leave-one-out cross-validation (LOOC) to avoid the randomness in splitting training and testing datasets [73]. With text features included, the best model achieves 83.56% accuracy (Kappa coefficient is 0.620), while the best accuracy is 79.41% without text features (Kappa coefficient is 0.496).

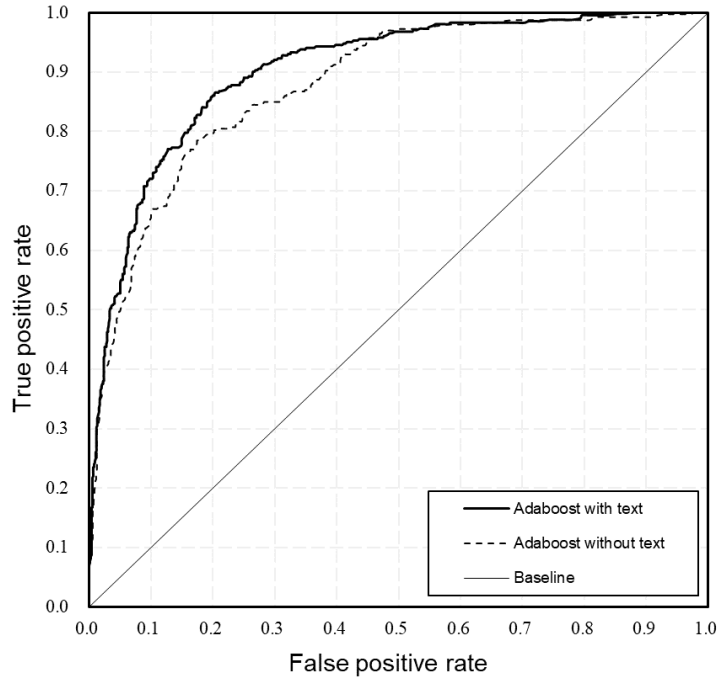

**Figure 5 The ROC curve of the different machine learning models**

The sample is approximately balanced, so the performance of the models should also be evaluated with respect to the sensitivity and specificity of prediction besides accuracy. In the context of this study, the *sensitivity* of the performance prediction model refers to the proportion of top performers that are correctly predicted. It is also known as true positive rate (TPR), i.e. truly positive as predicted. *Specificity* measures the proportion of non-top performers that are correctly predicted, and 1-*specificity* is also known as false positive rate (FPR), i.e. falsely predicted as positive. The direct outputs of a machine learning model are decision values for each test sample, and the threshold level should be set to give a prediction for each test sample. A lower threshold level makes it easier for truly positive samples to be predicted as positive and hence increases TPR. But this also makes more negative samples to be falsely predicted as positive, and hence increases FPR. A higher threshold level causes the contrary. Thus, there is a trade-off between increasing TPR and controlling FPR, and the predictive power of a model can be evaluated by the extent to which the increase in TPR can be achieved without too much increase in FPR. This can be directly evaluated by plotting FPR on the x-axis and TPR on the y-axis to construct a Receiver Operating Characteristic (ROC) curve. The more an ROC curve fills the Area Under the Curve (AUC), the higher the predictive power of a model. The ROC curves of the six models are plotted in Figure 5.

The Adaboost model with text features has the strongest predictive power with the ability to correctly identify more than 80% of the top performers and, at the same time, keeps the mistakes of incorrectly identifying non-top performers at less than 20%. Although interpreting the effects of predictors based on machine learning models is less straightforward, it is clear that the email indicators are highly predictive signals of individuals' work performance. The text features can also provide additional predictive power in identifying top performers.

## 5. Discussions

### 5.1 Predictive email communication indicators

According to Table 3, three network position indicators (betweenness, closeness and degree), one dynamic indicator (ego nudges), and four content indicators (influence, sentiment, complexity and emotionality) have significant associations with top performance. The machine learning model results in 4.2 further suggest that these indicators can act as the predictive signal of top performers with considerable predictive power.

Organizational email communication networks have been considered as highly constrained by predefined formal organization structures [22]. Thus, it remains to be examined whether email network position indicators can act as predictive signals of individuals' performance or are merely "shadows" of formal organization positions [41]. The empirical findings of this study support the relevance of the three centrality indicators in predicting work performance. The predictive power can be interpreted from three perspectives. First, central email network positions enable individuals to obtain a wide range of information, exercise control over information flows and have timely access to information [74]. These information advantages can translate to superior work performance. Second, voluntary informal interaction with coworkers beyond predefined formal organization structure is an important dimension of organization citizenship behavior, which contributes to high work performance [75]. Email provides an easy access to coworkers across the organization regardless of temporal or geographic restrictions and hence has a large potential to facilitate additional informal interactions [8, 37]. Such informal interactions are critical for accumulating social capital and completing collaborative tasks [76]. Third, central communication network positions imply more collaborative experience with coworkers, which according to the transactive memory theory, improves future collaboration efficiency [50, 77]. Individuals that are familiar with each other were found to outperform unacquainted individuals in collaborative tasks [78]. Therefore, individuals with central network positions are at an advantage in utilizing abundant transactive memory to achieve better performance.

Compared with questionnaire surveys based on respondents' memory, email archives provide a unique opportunity to obtain additional information on the contents of communication in real-world organizations. The four content indicators calculated from raw email texts are significant predictors of top performance. The more unique the information in an individual's email contents is, i.e. the higher his/her language complexity, the higher the individual's probability of being a top performer. This result is consistent with previous findings that email information diversity is positively associated with work performance [27]. The present study extends these findings in that the more coworkers adopt such diverse information in subsequent emails (higher influence indicator), the more likely the individual will be a top performer. Besides the informativeness of email contents, positive sentiment and low emotionality are associated with superior performance. This supports the notion that business communication should have less fluctuation in emotion, and expressing ideas with positive sentiment is beneficial [79].

*Ego nudges*, the average number of emails that need to be sent to an individual in order to get the individual's response, is the only dynamic indicator significantly associated with top performance. On the one hand, timely responses to emails are conducive to keeping coworkers in synchronization and coordinating collaborative tasks [8]. This is especially important for tasks with strong interdependencies [50]. On the other hand, as email overload becomes increasingly prevalent in modern organizations, individuals may find it hard to absorb information from huge amount of emails and give timely responses [33]. Beyond the time consumed by emails, emails may cause interruptions to individuals' work and lower work efficiency [31]. Therefore, an adequate email processing strategy is necessary to maintain the balance between giving timely responses and staying concentrated on tasks. Our empirical analysis results suggest that being responsive to more emails (*ego Nudges*) seems to be a more effective strategy than responding in shorter time (*ego ART*).

Besides, email text features also facilitate better prediction accuracy, suggesting that it is also important to select appropriate topics to discuss in emails. Taken together, these email indicators have considerable predictive power on work performance. Many existing studies imply that individuals' communication competence is a valuable skill in the digital communication environment compared to other communication channels [24]. However, the majority of existing studies on communication competence rely on self-report, which may miss many behavioral cues underpinning communication competence [80]. The email communication patterns of top performers identified in this study also act as the foundation for operationalizing the construct of email communication competence.

Compared to previous studies at the team-level [21], some enablers of superior team performance can be generalized to individual-level, such as high network centrality, high responsiveness and positive sentiment. However, not all team-level phenomena are supported at the individual-level. For example, relational leadership rotation (*betweenness oscillation*) and coworkers' high responsiveness (*alter ART* and *alter nudges*) are expected to promote team creativity and efficiency [53], but do not show significant association with individual performance. These results indicate that a top performer is not completely equivalent to an effective team collaborator, and imply different causal mechanisms to be further examined.

## 5.2 Versatile top performers

The predictive power of machine learning models suggests that there may exist complex interaction patterns among email indicators not captured by regression model. Therefore, we performed clustering analysis to further reveal top performers' email communication patterns. K-means clustering was employed to explore the underlying types of top performers based on their email communication indicators. The optimal number of clusters was determined by the "elbow" criterion [81], which balances the number of clusters (interpretability of clusters) and between-group variance maximization (explanatory power of clusters). The results suggest that three clusters of top performers emerged in the sample. In order to test the robustness of the clustering result, we also performed clustering analysis using the Gaussian mixture model based on the Expectation Maximization algorithm [82]. The Kappa inter-rater agreement coefficient between the two sets of clustering results is 0.954 ($p<0.001$) indicating strong consistency. Therefore, we consider the clustering results consistent and eligible for further interpretation. To visualize the clustering results, we conducted dimension reduction using principal component analysis (PCA) and plotted top performers on the coordinates of the first two principal components (Figure 5). The values of the three cluster centers in the dimension of each email indicator are presented using radar charts as the representative email communication profiles of the three kinds of top performers.

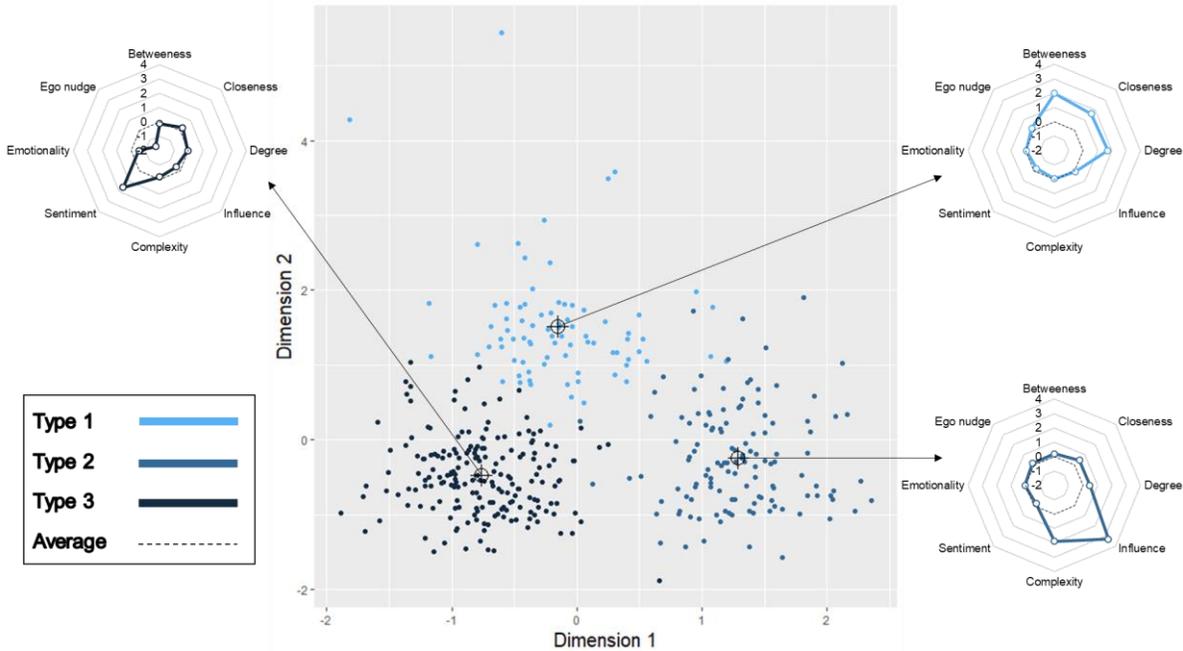

**Figure 6 The email communication profile of the three types of top performers**
**Note:** The values in the radar charts are measured in terms of standard deviations from the sample mean.

As shown in Figure 6, three kinds of top performers have rather distinct email communication profiles, which can be interpreted as: "networkers", 'influencers', and "positivists". Type 1 top performers locate at the center of the whole network with high network centrality values. Their superior social capitals in the email communication network enable them to act as the relational leaders that facilitate information diffusion in the organization. Type 2 top performers are only slightly higher in network centrality than average but have much higher influence and complexity indicator values. This suggests that their emails introduce more novel information (complexity) and tend to be followed by coworkers (influence). Thus, they can be regarded as the opinion leaders of the organization. Type 3 top performers stand out for their strongly positive sentiment, low emotion fluctuation and high responsiveness (low *ego nudges*) to coworkers' emails. They appear to be the emotional leaders that emit positive power and strengthen cohesiveness in the organization. The versatility of top performers' email communication profiles provides two fundamental implications for email communication competence.

First, not all top performers have the same pattern of email communication. There appears no unified set of optimal email communication pattern, and various patterns can be associated with top performance. This can be understood with respect to the fact that different roles, such as relational leaders ("networkers"), opinion leaders ("influencers") and emotional leaders ("positivists"), are needed to fulfill different organizational functions [83, 84]. This finding also implies the need to rethink the definition of email communication competence, which may be better operationalized as the combination of the strength in several communication dimensions instead of a unified construct.

Second, not all favorable communication behaviors (e.g. high responsiveness, positive sentiment) reside in one kind of top performers. This, on the one hand, reflects the inherent trade-off between the focus on different communication behavior given an individual's limited time and energy. For example, it may be hard for an individual to be highly responsive (type 3) and at the same time keep introducing novel and influential ideas (type 2). On the other hand, as shown in Figure 6, the three kinds of top performers are rather specialized in a few dimensions of email indicators and are average (or even below average) in other dimensions. This can also be viewed as top performers' adaptation to organizational needs instead of being dominant in every respect.

## 6. Conclusion

This study reveals the email communication indicators that are predictive of individuals' work performance. The panel regression models provide interpretable results that top performance is associated with central network position, positive sentiment, low emotionality, high complexity, more adoption of influential words by coworkers and high responsiveness in email communication. The machine learning models that allow more complex interactions among independent variables indicate the high predictive power of email indicators with the best model achieving over 80% accuracy. The cluster analysis results further reveal that the top performers can be generally classified into three types that have advantages in different dimensions.

For theory development, the findings provide implications for operationalizing the construct of email communication competence with subjective measures from real-world email data. The variation in communication style of top performers further implies that email communication competence might be better defined as a combination of several supportive factors instead of a unified construct. For management practices, the identified email indicators can be used to suggest improvements in email communication skill development training. Furthermore, as fine-grained email communication data becomes increasingly available, the analyses performed in this study can also be replicated in different organizations to understand individuals' contributions to organization communication.

The implications of this study should be viewed with respect to the following limitations, which leave room for improvements in future studies. First, the data are from only one company. Although the fact that the company operates internationally alleviates this limitation to some extent, caution should still be taken when generalizing the findings to other organizational settings. Future studies to test the robustness of the findings in a broader range of cultural and organization environments are needed. Second, because of privacy restrictions we cannot directly observe the email contents, which may provide additional information. Third, the findings of this study cannot support causal interpretation. Future research may use causal inference techniques, such as natural experiments, to precisely estimate the effects of influential email indicators.

# Appendix

**Appendix Table 1 Email communication indicators in existing studies**

| | Indicators | Definition and calculation method | Related research[1] | Outcome variables |
|---|---|---|---|---|
| Network position | Betweenness centrality | The extent to which the person acts as an information hub and controls information flow | Ahuja et al. [41] | Individual performance |
| | | | Aral and Van Alstyne [42] | Worker productivity |
| | | | Trier [74] | Group evolution |
| | | | Hossain and Wu [85] | Coordination behavior |
| | | | Wen and Qiang [86] | Information sharing |
| | | | Kidane and Gloor [53] | Team performance |
| | | | Gloor et al. [13] | Job turnover |
| | Closeness centrality | The extent of the person's affinity with co-workers | Ahuja et al. [41] | Individual performance |
| | | | Aral and Van Alstyne [42] | Worker productivity |
| | | | Trier [74] | Group evolution |
| | | | Hossain and Wu [85] | Coordination behavior |
| | | | Gloor et al. [13] | Job turnover |
| | | | Wen and Qiang [86] | Information sharing |
| | Degree centrality | Number of co-workers the person communicates | Ahuja et al. [41] | Individual |

| | | | with. | performance |
|---|---|---|---|---|
| | | | Aral and Van Alstyne [42] | Worker productivity |
| | | | Trier [74] | Group evolution |
| | | | Hossain and Wu [85] | Coordination behavior |
| | | | Aral et al. [36] | Individual performance |
| | | | Wen and Qiang [86] | Knowledge sharing |
| | | | Gloor et al. [13] | Job turnover |
| Contribution | Messages sent | Number of emails the person sent | Wasiak et al. [18] | Team performance |
| | | | Jarvenpaa et al. [35] | Team performance |
| | | | Kanawattanachai and Yoo [50] | Team performance |
| | | | Gloor et al. [13] | Job turnover |
| | | | Aral et al. [36] | Worker productivity |
| | | | Aral and Van Alstyne [42] | Individual performance |

**Appendix Table 1 Email communication indicators in existing studies (continued)**

| | Indicators | Definition and calculation method | Related research[1] | Outcome variables |
|---|---|---|---|---|
| Contribution | Balance of contribution | The extent to which the person's communication is balanced in terms of emails sent and received | Eckmann et al. [54] | Group cohesion |
| | | | Lim and Teo [39] | Work attitude |
| Content | Sentiment | Positivity and negativity of communication | Wasiak et al. [18] | Team performance |

|  |  |  | Gloor et al. [13] | Job turnover |
|  |  |  | Gloor [57] | Communication behavior |
|  | Emotionality | Deviation from neutral sentiment | Francis et al. [40] | Communication behavior |
|  |  |  | Gloor et al. [13] | Job turnover |
|  |  |  | Lim and Teo [39] | Work attitude |
|  | Complexity | The complexity of a person's email contents | Erhardt et al. [4] | Team learning |
|  |  |  | Kanawattanachai and Yoo [50] | Team performance |
|  |  |  | Gloor et al. [13] | Job turnover |
|  |  |  | Aral et al. [36] | Individual performance |
|  | Influence | The influence of a person's email contents on co-workers' email contents | Gloor et al. [13] | Job turnover |
|  |  |  | Iribarren and Moro [63] | Information sharing |
|  |  |  | Gloor [57] | Communication behavior |
| Dynamics | Ego responsiveness | The person's responsiveness to emails | Erhardt et al. [4] | Team learning |
|  |  |  | Gloor et al. [13] | Job turnover |
|  |  |  | Eckmann et al. [54] | Group cohesion |
|  | Alter responsiveness | The responsiveness of the person's co-workers | Eckmann et al. [54] | Group cohesion |
|  |  |  | Gloor et al. [13] | Job turnover |
|  | Network position rotation | The extent to which the person frequently changes network position from central to peripheral, and back | Gloor et al. [13] | Job turnover |
|  |  |  | Trier [74] | Group evolution |
|  |  |  | Gloor [57] | Communication behavior |

[1] The references in this table are restricted to the studies using emails in real-world work settings as data source. Therefore, some studies on email communication using subjective measures (e.g. [3]) are not included.